\documentclass[a4paper,11pt]{article}
\usepackage{jinstpub} % for details on the use of the package, please see the JINST-author-manual
\title{\boldmath The ultra-low material budget GEM-based TPC for tracking with VMM3a readout}

\author[a,*]{F. Garcia}
\author[c,i,*]{, K.J. Flöthner}
\author[b]{, A. Amato}
\author[b]{, S. Biswas}
\author[c]{, F.M. Brunbauer}
\author[b]{, M. W. Heiss}
\author[b]{, G. Janka}
\author[c]{, D. Janssens}
\author[c,g]{, M. Lisowska}
\author[c,f]{, M. Meurer} 
\author[c,d]{, H. Muller}
\author[b,e]{, B. Banto Oberhauser}
\author[c]{, E. Oliveri}
\author[c]{, G. Orlandini}
\author[c,h]{, D. Pfeiffer}
\author[b]{, T. Prokscha}
\author[c]{, L. Ropelewski}
\author[c]{, L. Scharenberg}
\author[c,h]{, J. Samarati}
\author[c]{, F. Sauli}
\author[c]{, M. van Stenis}
\author[c]{, R. Veenhof}
\author[b,e]{, B. Zeh}
\author[b]{, X. Zhao}

\affiliation[a]{Helsinki Institute of Physics, University of Helsinki, Gustaf Hällströmin katu 2a, Helsinki, FI-00014, Finland}
\affiliation[b]{PSI Center for Neutron and Muon Sciences CNM, 5232 Villigen PSI, Switzerland}
\affiliation[c]{European Organization for Nuclear Research (CERN),1211 Geneva 23, Switzerland}
\affiliation[d]{Physikalisches Institut, University of Bonn,Nußallee 12, 53115 Bonn,Germany}
\affiliation[e]{ETH Zürich, Swiss Federal Institute For Technology Institute for Particle Physics and Astrophysics (IPA),Otto-Stern-Weg 5, 8093 Zürich,Switzerland}
\affiliation[f]{Ludwig Maximilian University of Munich, Am Coulombwall 1, 85748, Garching, Germany}
\affiliation[g]{Université Paris-Saclay, F-91191 Gif-sur-Yvette, France}
\affiliation[h]{European Spallation Source ERIC (ESS), Box 176, SE-221 00 Lund, Sweden}
\affiliation[i]{Helmholtz-Institut f\"{u}r Strahlen- und Kernphysik, University of Bonn, Nu\ss{}allee 14-16, 53115 Bonn, Germany}

\emailAdd{Francisco.Garcia@helsinki.fi}
\emailAdd{karl.jonathan.floethner@cern.ch}

\abstract{
The Gaseous Electron Multiplier-based Time Projection Chamber (GEM-TPC) in TWIN configuration for particle tracking has been consolidated after extensive investigations in different facilities to study its tracking performance.
The most attractive feature of this detector is its ultra-low material budget, which is 0.28\% X/X$_0$ and can be further reduced by decreasing the thickness of the gas traversed by the incident particles. Thus, it provides excellent position reconstruction and reduced multi-scattering.

This detector consists of two GEM-TPCs with drift fields in opposite directions, achieved by rotating one 180 degrees in the middle plane with respect to the other. These two GEM-TPCs share the same gas volume, i.e., inside a single vessel. This configuration is called a TWIN configuration.
The results presented in this work were measured using the newly integrated VMM3a—SRS readout electronics, an important milestone in improving overall performance and capabilities.

In 2024, this detector was tested at the H4 beamline of the SPS at CERN, using muons and pions and with different gas mixtures like, for instance: Ar/CO$_2$ (70/30 \%), He/CO$_2$ (70/30 \%) and He/CO$_2$ (90/10 \%). The helium-based mixtures were used to commission the detector to track low momenta muons required in the PSI muon-induced X-ray emission (MIXE) experiment.

The results obtained from these measurements, a brief discussion of the methodology used for the data analysis, and a comparison of the spatial resolution for different gas mixtures will be presented.
}


\arxivnumber{6128024} % Only if you have one

\begin{document}
\maketitle
\flushbottom

\section{Introduction}
\label{sec:intro}
\begin{figure}[h!]
\centering % \begin{center}/\end{center} takes some additional vertical space
\raisebox{-0.3\height}{\includegraphics[width=0.39\textwidth]{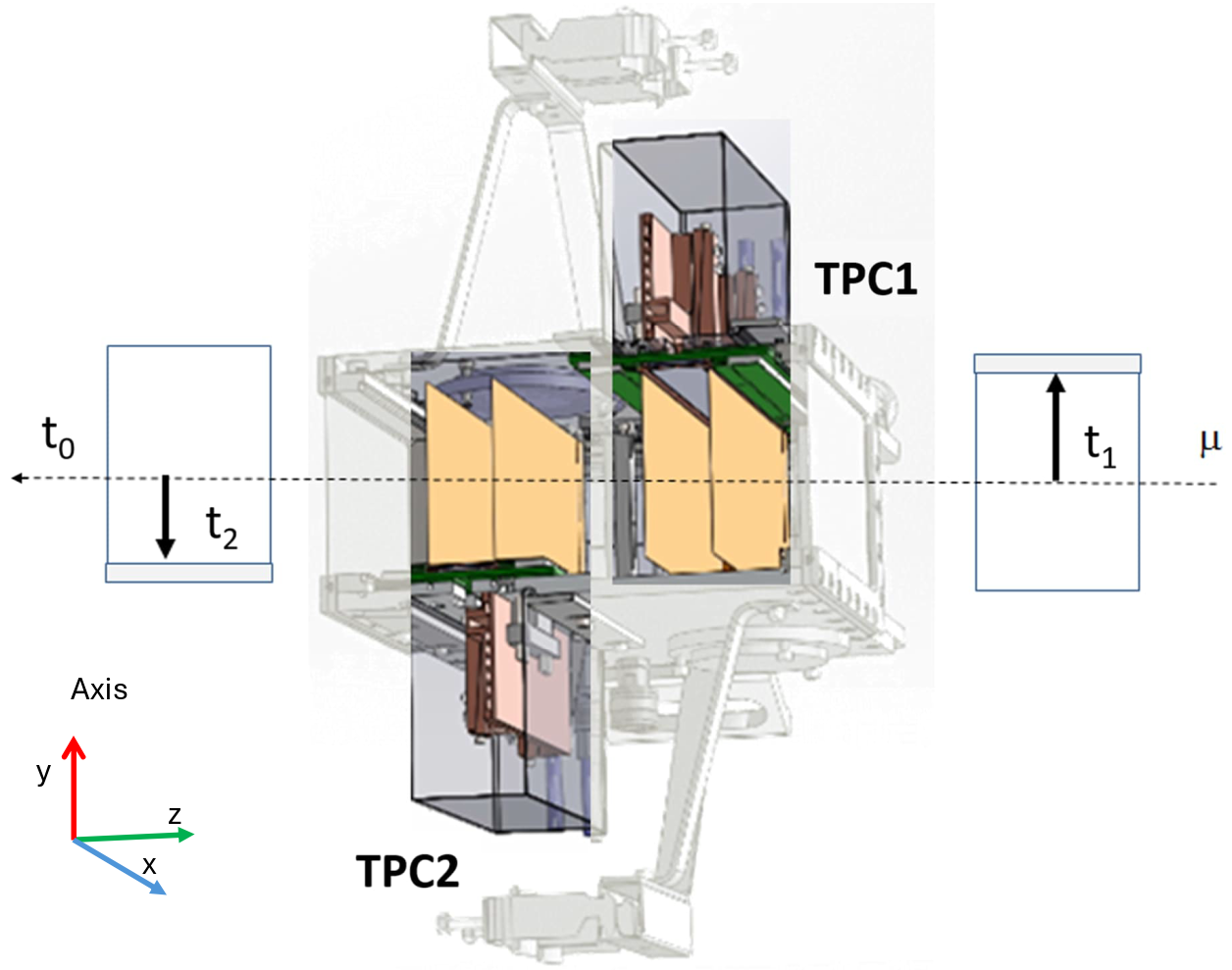}}
\raisebox{-0.3\height}{\includegraphics[width=0.39\textwidth]{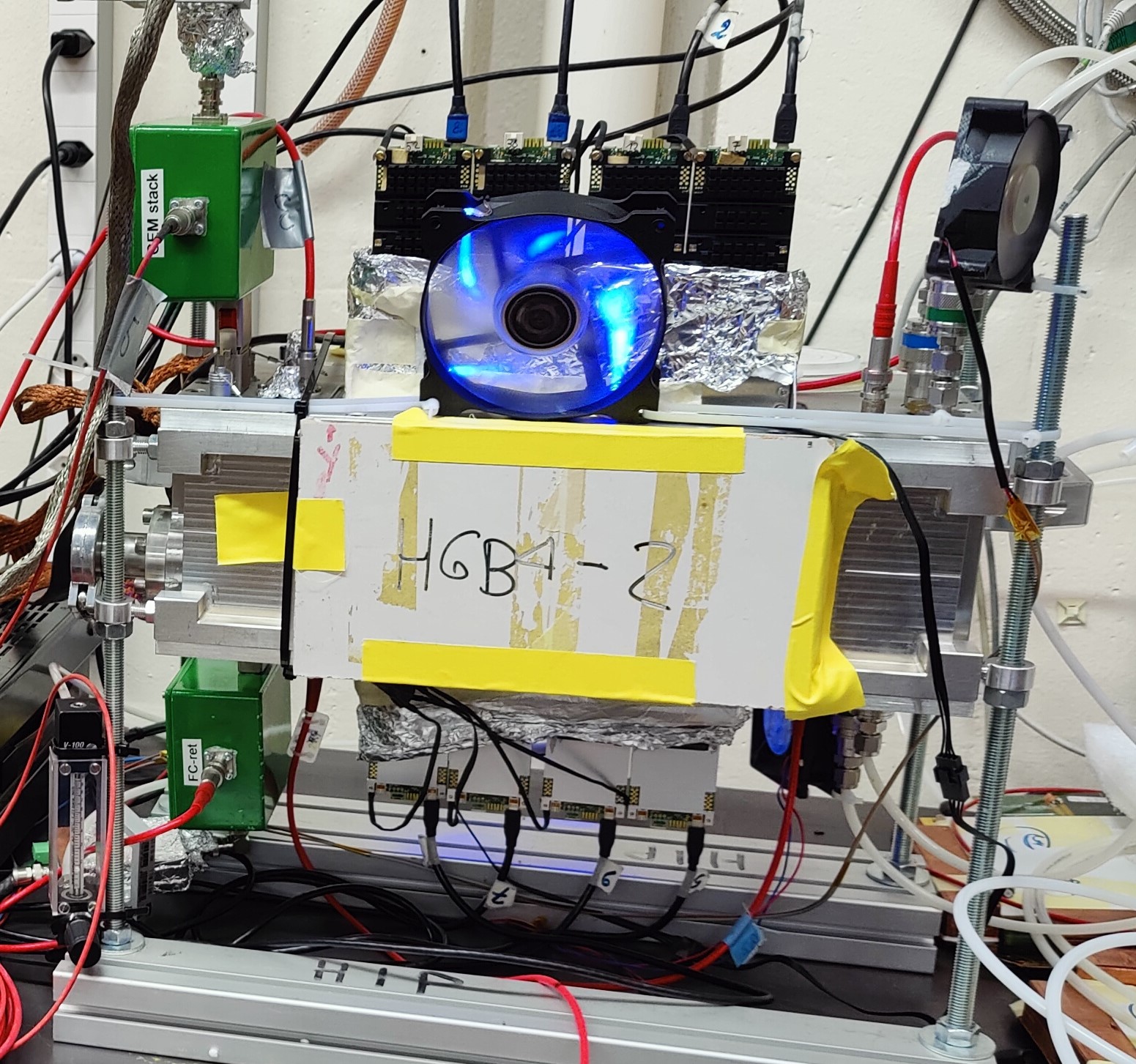}}
%"\includegraphics" from the "graphics" permits to crop (trim+clip)
% and rotate (angle) and image (and much more)
\caption{\label{fig: TWIN} The GEM-TPC in twin configuration schematics (left) illustrating the principle of operation. It has two GEM-TPCs, with TPC1 facing the drift to the top and TPC2 to the bottom, inside a single vessel. The HGB4-2 after integrating the VMM3a-SRS during commissioning at the GDD laboratory at CERN (right).}
\end{figure}

Twin-time projection chambers (TWIN TPC) were introduced to track heavy ions in the Super-FRS to cope with an increased particle rate. The conventional TPCs\cite{Janik2009} fitted into the twin TPC used for the amplification stage, a set of four single proportional counters. The strip is read out with delay lines. Even though this concept represented a good choice to cope with high particle rates, the signal length and the low speed of the delay line were the limiting factors.
Therefore, to overcome those limitations, the amplification stage was replaced by a Gas Electron Multiplier (GEM)\cite{Sauli1997531} and the readout by fast and high-density readout electronics, resulting in the GEM-TPC in TWIN configuration concept~\cite{Garcia:183901, Garcia2018}.
The second prototype of the four generation of these detectors, named HGB4-2, which stands for Helsinki-GSI-Bratislava, is shown in Fig.~\ref{fig: TWIN} and is the subject of the investigation for this work.

This detector has two GEM-TPCs inside a single vessel with the same gas volume. One is rotated 180 degrees in the middle plane with respect to the other. Upstream, the TPC1 has the amplification stage and readout electronics on top, and the second, TPC2, on the bottom. This detector has an ultra-low material budget, which is 0.28\% X/X$_0$, and can be further reduced by decreasing the thickness of the gas traversed by the incident particles.

In this configuration, the electric fields of the two field cages are in opposite directions. As a result, the electrons generated by the traversing particles drift in TPC1 toward the top and in TPC2 to the bottom.

The readout plane consists of a single GEM-TPC is 512 strips, with a width of 300$\mu$m and, a pitch of 400$\mu$m. The total number of strips for the whole detector is 1024 strips. The strips are parallel to the incoming beam direction on both TPCs.

The newly integrated continuous readout electronics VMM3a-SRS\cite{LUPBERGER201891, PFEIFFER2022166548} collects the signals of each strip of a single event, digitizes the amplitude, and sets the arrival time stamp. Then, a clustering procedure groups the fired strips of a single event. 

Recording the signals from a trigger scintillator, which sets the event's starting time or the t0, makes it possible to reconstruct the coordinates on the vertical plane of a track on both GEM-TPCs. In the horizontal plane, the coordinate is determined by the projection of the cluster in the strip plane. Hence, the two coordinates of the track are found, and the position of the tracks is reconstructed.

Section 2 briefly describes the methodology used for the data analysis. Section 3 discusses spatial resolution, the main observable investigated in this work, for different gas mixtures.

\section{Experimental Setup}

The test beam was carried out at the H4 beamline\cite{H4Beamline} of the SPS at CERN during the DRD1 test beam in 2024. The main characteristics of the 150GeV/c beam momenta were a wide muon beam and a very narrow one, for pions. The flux was up to 200k particles per spill.
The HGB4-2 was located on a mobile platform in front of the RD51 reference tracker\cite{Scharenberg2023} to overlap both sensitive areas and perform a position scan in the entrance window, which is 200 mm in the horizontal plane and 100 mm in the vertical plane for the HGB4-2 and 100 mm in both planes for the RD51 reference tracker.
The readout plane of the HGB4-2 is made of strips oriented parallel to the incoming particle beam in 1D, while for the reference tracker, each station has strips in 2D geometry.

\begin{figure}[h!]
\centering % \begin{center}/\end{center} takes some additional vertical space
\includegraphics[width=0.52\textwidth]{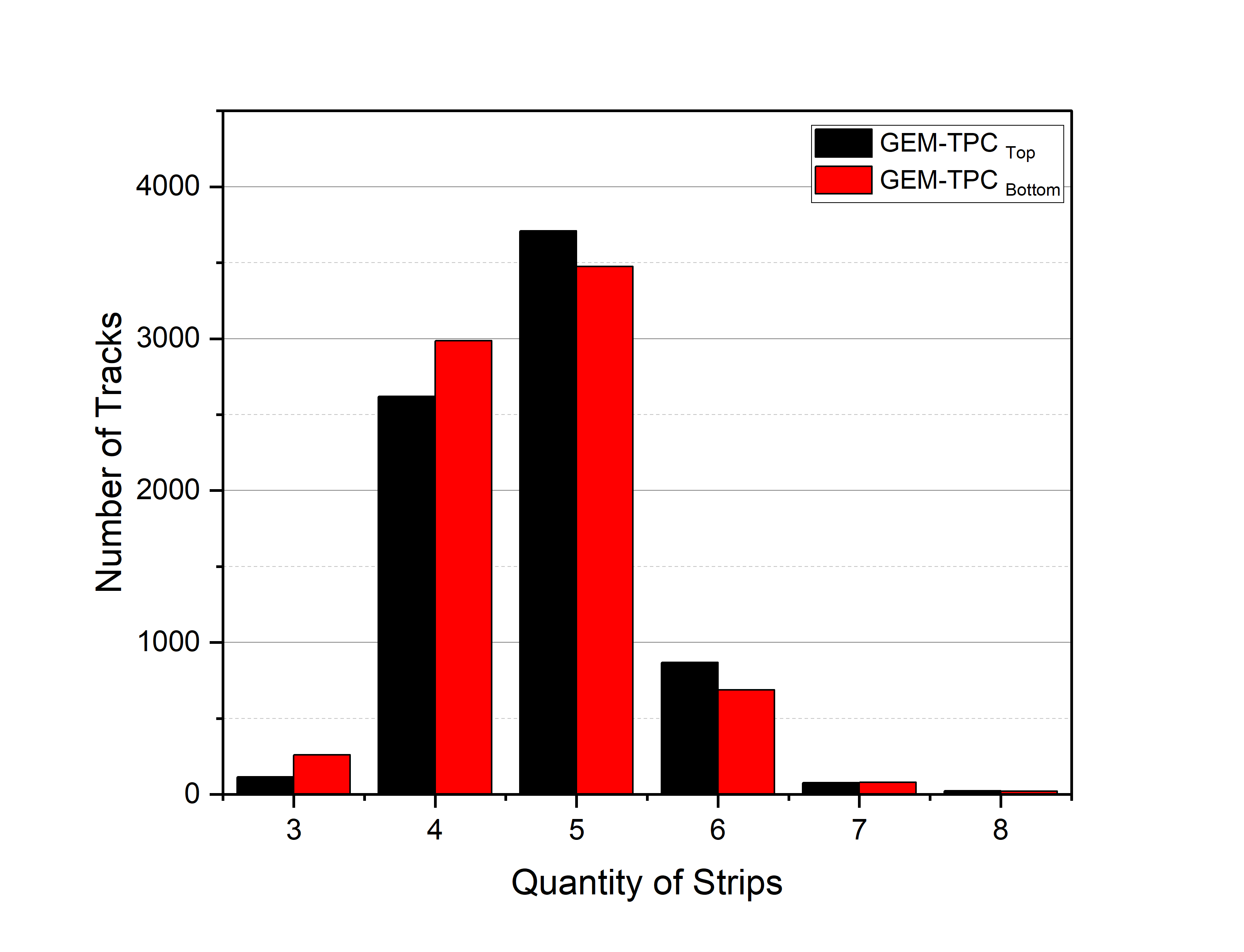}
% "\includegraphics" from the "graphics" permits to crop (trim+clip)
% and rotate (angle) and image (and much more)
\caption{\label{fig:not0profile} Cluster Strip multiplicity of the HGB4-2, for both GEM-TPCs. The mean occupancy is similar for the GEM-TPC Top (in black) and the GEM-TPC Bottom (in red).}
\end{figure}

For this campaign, the gas mixtures used were Ar/CO$_2$ (70/30 \%), He/CO$_2$ (70/30 \%), and He/CO$_2$ (90/10 \%). The helium-based gas mixtures were used to commission the HGB4-2 for the PSI MIXE experiment\cite{app12052541}.
The signals were acquired using the VMM3a-based RD51 scalable readout system.
The Front-End chip is the VMM3a ASIC~\cite{Iakovidis2020}, which provides the peak amplitude and time at peak for signals crossing a threshold level.

The VMM3a Slow Control and Calibration Software~\cite{VMMsc}, developed by the RD51 collaboration, was used to control the electronics.
For collecting and recording data, tcpdump~\cite{tcpDump} was used. For clustering and reconstruction, vmm-sdat~\cite{vmmsdat} was used. A Python-based analysis was developed for tracking performance analysis.

The alignment procedure ensured a good overlap of the detection areas of both systems. The drift velocity was calibrated, yielding a 1 - 2\% difference between the GEM-TPC top and bottom.
The signal amplitude distribution was measured to assess the level of equalization of their response and to identify the amount of saturated channels. The cluster strip multiplicity obtained agrees with those expected from the computed values of traverse diffusion.
Fig. 2 shows the cluster strip multiplicity for the Ar/CO$_2$ (70/30 \%) gas mixture. It can be seen that the two distributions are very similar and the mean occupancy of five strips agrees with the expected traversal diffusion, indicating a very similar response of both GEM-TPCs, to the incoming beam.

\section{Spatial resolution for different gas mixtures}

The main observable under investigation was the spatial resolution of the HGB4-2, GEM-TPC in twin configuration as a function of the electric field strength of the field cage, as mentioned above. The second aspect of interest is the long-term stability at a large gas gain. The muons and pions deposit an energy close to that of a proton in the MIP region traversing the detector. Therefore, these measurements allow us to assess the response to protons.

\begin{figure}[h!]
\centering % \begin{center}/\end{center} takes some additional vertical space
\includegraphics[width=0.64\textwidth]{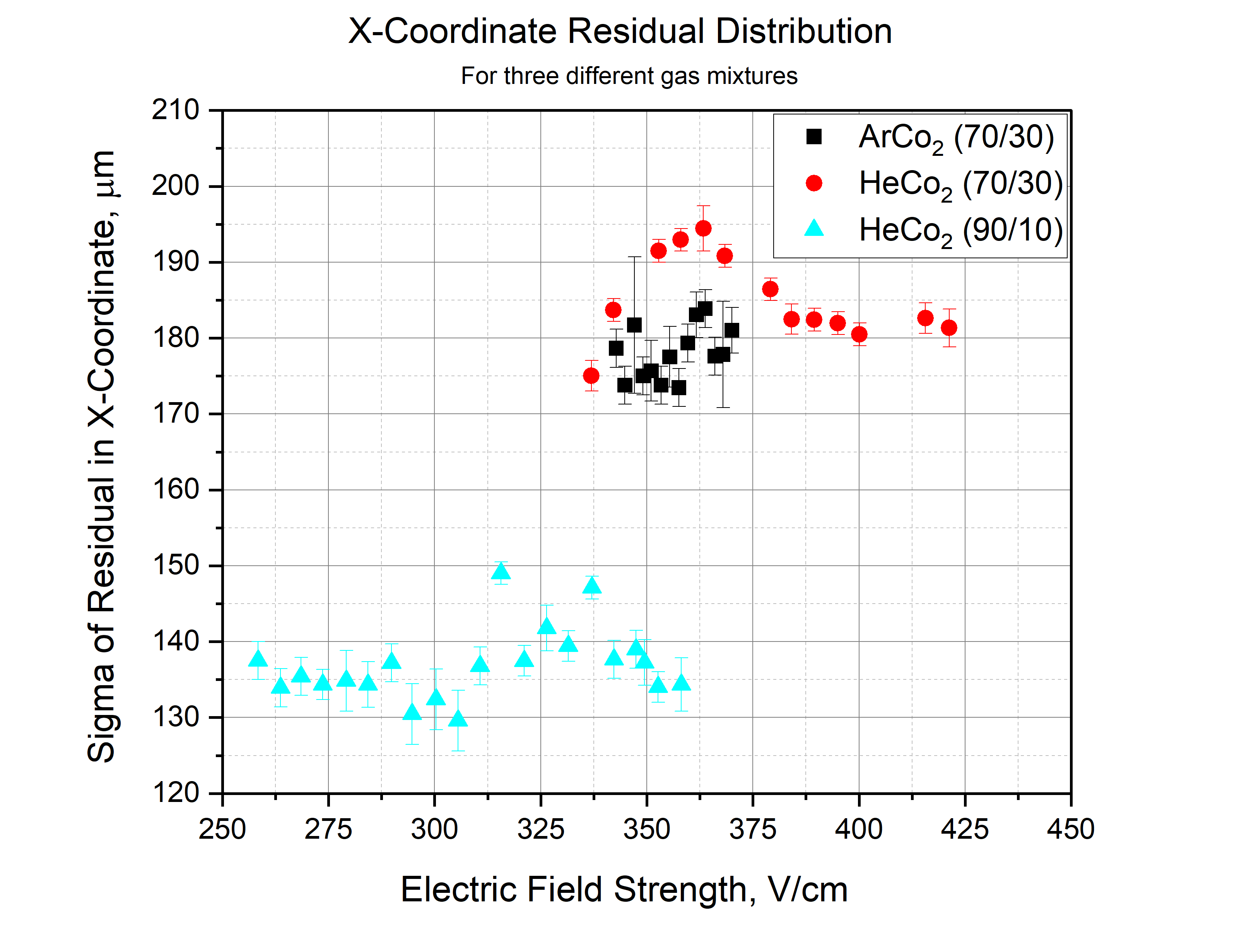}
\includegraphics[width=0.64\textwidth]{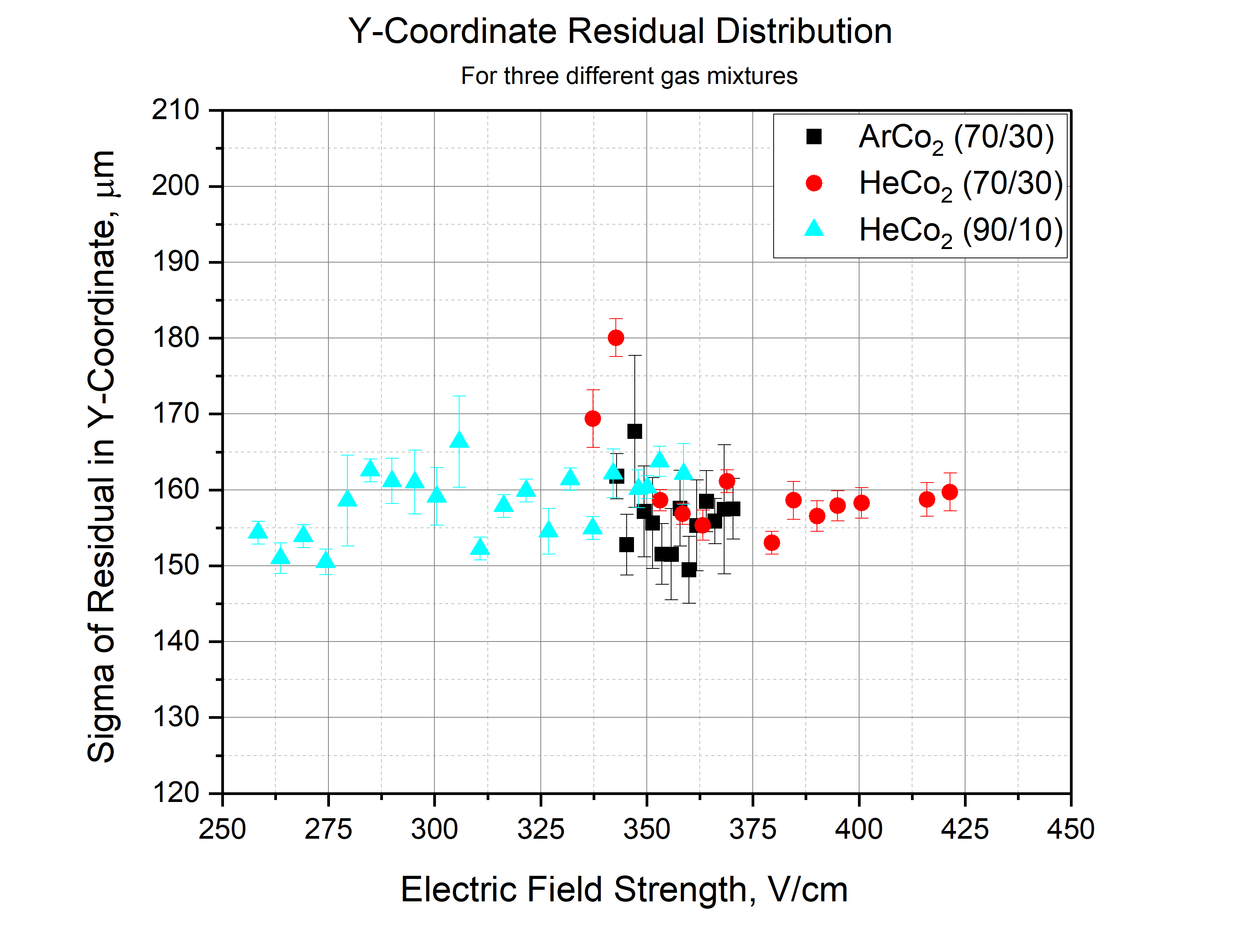}
% "\includegraphics" from the "graphics" permits to crop (trim+clip)
% and rotate (angle) and image (and much more)
\caption{\label{fig:t0proove} Residual Distributions of the track reconstructed in HGB4-2 for muons; on the top for the x-coordinate and on the bottom for the y-coordinate. It can be seen that while in the x-axis, there is a notable difference between Ar/CO$_2$ (70/30 \%) and He/CO$_2$ (90/10 \%) with 180$\mu$m versus 135$\mu$m. For the y-axis, the resolution is quite constant at about 155$\mu$m.}
\end{figure}

The operation was very stable, with no destructive discharges during three weeks of continuous operation. The system performed well and collected data uninterruptedly, without major incidents, for long periods at a particle rate of up to 50 kHz. These aspects were an essential contribution to consolidating the concept and provided the ground for its evolution as a tracking device with the advantage of a very low material budget.

Fig. 3 shows the spatial resolution measured with a muon beam for three gas mixtures. It can be seen that in the horizontal plane, i.e., x-coordinate, the He/CO$_2$ (90/10 \%)  gives the best resolution of about 135$\mu$m while for both He/CO$_2$ (70/30 \%) and Ar/CO$_2$ (70/30 \%) the resolution is around 180$\mu$m. These results agree with the expected values from the traverse diffusion.

For the vertical plane, the coordinate y for all gas mixtures is obtained with a resolution of 155$\mu$m. The reasons behind such behavior are not yet clear. Therefore, more effort will be devoted to finding a better algorithm for timing the clusters.

The next step is to integrate a pixelated readout plane\cite{scharenberg2024mpgdsembeddedpixelasics}, which allows angular correction of the incoming tracks and improves the precision of the track reconstruction.

\acknowledgments

We thank the RD51 test beam workgroup for the organization of this campaign. Many thanks in GSI to C. Kaya, J. Kunkel, B. Voss, and H. Risch for the design and field cages, and to Christian Schmidt for facilitating the assembly infrastructure. R. Turpeinen produced the flanges and window frames at the Helsinki Institute of Physics Mechanical Workshop. J. Heino for quality assurance and framing. Thanks to S. Rinta-Antila for providing the detector housing.
The authors thank R. de Oliveira and the CERN Micro Pattern Technology (MPT) Workshop for supporting and producing the GEM foils and frames. This work is sponsored by the Wolfgang Gentner Programme of the German Federal Ministry of Education and Research (grant no. 13E18CHA). The ETH Zurich and SNSF Grant Nos. 169133, 186181, 186158, 197346, and 219485 (Switzerland). The CERN Strategic Programme on Technologies for Future Experiments (\url{https://ep-rnd.web.cern.ch/}) and the European Union's Horizon 2020 Research and Innovation programme under Grant Agreement No 101004761.

%\paragraph{Note added.} This is also a good position for notes added
%after the paper has been written.

% Bibliography

%% [A] Recommended: using JHEP.bst file
%% \bibliographystyle{JHEP}
%% \bibliography{biblio.bib}

%% or
%% [B] Manual formatting (see below)
%% (i) We suggest to always provide author, title and journal data or doi:
%% in short all the informations that clearly identify a document.
%% (ii) please avoid comments such as "For a review'', "For some examples",
%% "and references therein" or move them in the text. In general, please leave only references in the bibliography and move all
%% accessory text in footnotes.
%% (iii) Also, please have only one work for each \bibitem.

\newpage
 \bibliographystyle{JHEP}
 \bibliography{biblio}
%
%\begin{thebibliography}{99}

%\bibitem{a}
%Author,
%\emph{Title},
%\emph{J. Abbrev.} {\bf vol} (year) pg.

%\bibitem{b}
%Author,
%\emph{Title},
%arxiv:1234.5678.

%\bibitem{c}
%Author,
%\emph{Title},
%Publisher (year).

%\end{thebibliography}

\end{document}